\documentclass[
  reprint,
  aps,
  prresearch,
  superscriptaddress
]{revtex4-2}

\usepackage{amsmath} 
\usepackage{placeins}

\usepackage[braket, qm]{qcircuit}
\usepackage{tikz}
\usetikzlibrary{decorations.pathreplacing}

\usepackage{algorithm}
\usepackage{algpseudocode}

\begin{document}
\title{Applicability and Limitations of Quantum
        Circuit Cutting in Classical State-Vector Simulation}

\author{Mitsuhiro Matsumoto}
\email{mitsuhiro.matsumoto@pwc.com}
\affiliation{Technology Laboratory, PwC Consulting LLC,
            Japan}
\author{Shinichiro Sanji}
\affiliation{Technology Laboratory, PwC Consulting LLC,
            Japan}
\author{Takahiko Satoh}
\affiliation{Faculty of Science and Technology, Keio University, 
            Japan}

\begin{abstract}
Circuit cutting partitions a large quantum circuit into
smaller subcircuits that can be executed independently
and recombined by classical post-processing.
In classical state-vector simulation with full-state
reconstruction, the runtime is governed by a trade-off between
reduced subcircuit size and the overheads of
exponentially many subcircuits and full-state
reconstruction.
For equal partitioning, we derive threshold
conditions on the number of cuts below which
cutting reduces the wall-clock time.
State-vector experiments validate the predicted speedup
boundary up to 24 qubits, and a runtime breakdown up
to 30 qubits identifies crossovers at $q \approx 18$
and $q \approx 22$ where merging overtakes
first preprocessing and then subcircuit simulation.
As a practical guideline, we show that under a
10-minute wall-clock budget, two-way cutting extends
the maximum feasible qubit count by 4 to 6~qubits
relative to simulation without cutting.
\end{abstract}

\maketitle

\section{Introduction}
\label{sec:intro}
Quantum computing holds the potential to exponentially
outperform classical computing for specific computational
problems~\cite{Shalf15}.
However, current Noisy Intermediate-Scale Quantum (NISQ)
devices remain severely constrained by the number and
quality of available qubits~\cite{Preskill18}.
To overcome these limitations, circuit cutting, a technique
that partitions large quantum circuits into smaller
subcircuits for independent execution, has emerged 
as an approach to extend the effective circuit size beyond near-term device limits~\cite{bravyi2016trading,Peng20,Mitarai21}.

Existing research on circuit cutting has
focused on minimizing the overhead associated with
partitioning~\cite{Brenner23,Harada23}.
On the classical simulation side, hybrid
Schr\"{o}dinger--Feynman methods~\cite{markov2018quantum} and
tensor-network-based approaches~\cite{Pan2022sycamore, Pednault17}
exploit circuit partitioning to push simulation beyond naive
memory limits, and the fundamental trade-off between
partition size and the exponential growth of subcircuit
count is well recognized.
Classical simulation remains indispensable for verification, benchmarking, and workflow-level performance studies of quantum algorithms, especially when end-to-end runtime is the quantity of interest.
However, under concrete execution pipelines for classical state-vector simulation with full-state reconstruction, the cut count at which end-to-end wall-clock overhead outweighs the reduced subcircuit size has received limited attention.
A closed-form guideline for the cut threshold would help allocate finite classical resources and clarify when cutting is practically beneficial in wall-clock time.

Relatedly, circuit cutting has been studied as a workflow that distributes a large circuit into subcircuits and recombines them by classical post-processing~\cite{Tang2021CutQC}. 
The practical classical simulation frontier is likewise shaped by high-performance state-vector simulators ~\cite{Jones2019QuEST,Villalonga2018FlexibleSimulator,Suzuki2021Qulacs}, tensor-network backends~\cite{Nguyen2021TNQVM}, and large-scale partitioning demonstrations~\cite{Arute2019QuantumSupremacy}. 
These developments motivate an end-to-end wall-clock analysis of circuit cutting, which is the focus of this paper.

This work focuses on classical state-vector simulation
with full-state reconstruction, arising in
benchmarking, debugging, and verification of quantum
devices.
Cutting reduces the subcircuit size but exponentially
increases the number of subcircuits, and in this setting
the results must be combined into the full
$2^q$-dimensional state vector at an additional $O(2^q)$ merging cost.
Judging whether cutting is beneficial therefore requires
an end-to-end wall-clock analysis covering all stages of
the cutting workflow.
To our knowledge, a quantitative wall-clock analysis under this specific setting has not been systematically reported.

We analyze the wall-clock efficiency of circuit cutting
in classical state-vector simulation, presenting three
main contributions:

First, we decompose the total cutting time into
preprocessing, subcircuit simulation, and merging, and
analyze their relative scaling. This reveals two
bottleneck crossovers as qubit count increases: merging
surpasses preprocessing and surpasses subcircuit
simulation. 
In our implementation, these crossovers appear at intermediate qubit counts (around $q\approx 18$ and $q\approx 22$ in our setup).
Second, from this cost model we derive conditions on
the maximum total cut count $c_{\mathrm{total}}$ for
which cutting yields a net speedup under equal
partitioning into $n$ segments.
The threshold takes the form
$c_{\mathrm{total}} \propto q$ and is approximately
depth-independent.
State-vector experiments up to 24~qubits confirm the
predicted threshold.

Third, under a 10-minute wall-clock budget, two-way
cutting extends the maximum feasible qubit count by 4
to 6~qubits compared to simulation without cutting,
with the gain increasing at larger depths.

We begin by reviewing circuit cutting methods and
their computational characteristics in
Sec.~\ref{sec:background}.
We introduce the execution time model and its
regime structure in Sec.~\ref{sec:Theory}, and
derive cut-threshold conditions for equal
partitioning into $n$ segments in
Sec.~\ref{sec:threshold}.
We present numerical results and confirm the predicted
trends using classical state-vector simulations in
Sec.~\ref{sec:NV}.
Finally, we summarize our findings and conclude
in Sec.~\ref{sec:conclusion}.
\section{Background}
\label{sec:background}

\subsection{Circuit cutting}
\label{sec:cutting}

Circuit cutting decomposes a large circuit into
an ensemble of smaller subcircuits whose outcomes are
recombined by classical post-processing.
In practice, two representative approaches are
(i)~\emph{wire cutting} (cutting ``wires'' by inserting
measure-and-prepare channels) and
(ii)~\emph{gate cutting} (replacing non-local entangling
gates across partitions with decomposed local operations).
This work focuses on gate cutting in the classical
state-vector setting, whose overhead structure
differs qualitatively from the hardware setting.

\subsubsection{Quantum hardware setting}
\label{sec:cutting_hardware}

On quantum hardware, circuit cutting incurs a
sampling overhead that grows exponentially with the
number of cut locations.
For gate cutting, each non-local gate is replaced by a
quasiprobability decomposition (QPD) into local
channels~\cite{Mitarai21}.
Optimal sampling overheads for cutting multiple
two-qubit unitaries have been
characterized~\cite{schmitt2025cutting}.
For wire cutting, qubit wires are severed by
inserting measure-and-prepare
channels~\cite{Peng20}.
The sampling cost scales as
$\widetilde{\mathcal{O}}(4^{m}/\epsilon^{2})$
for cutting $m$~parallel wires
(up to polylog factors),
with an information-theoretic lower bound of
$\Omega(2^{m}/\epsilon^{2})$~\cite{Lowe2023fastquantumcircuit}.
Recent works study wire-cutting schemes that
achieve optimal or near-optimal overheads under
different operational constraints such as ancilla
usage~\cite{Brenner23,Harada23}.

\subsubsection{Classical state-vector setting}
\label{sec:cutting_classical}

When subcircuits are simulated classically via
state-vector simulation, a different decomposition
is available.
Any two-qubit unitary $U$ admits an
operator-Schmidt
decomposition~\cite{balakrishnan2011operator}
\begin{equation}
  U = \sum_{\ell=1}^{k} s_\ell \,
      (A_\ell \otimes B_\ell),
  \label{eq:schmidt}
\end{equation}
where $A_\ell$ and $B_\ell$ are single-qubit operators,
$s_\ell$ are complex coefficients, and
$k$ is the Schmidt rank of~$U$.
For two-qubit gates, the Schmidt rank takes only
the values $k=2$ (controlled unitaries such as
CZ and CX) or $k=4$ (all other non-local gates
such as SWAP)~\cite{balakrishnan2011operator}.

Because the classical simulator has direct access
to the full state vector, the linearity of
Eq.~(\ref{eq:schmidt}) can be exploited directly:
each term produces an independent subcircuit
state vector, and the results are summed as
complex vectors.
This yields $k$~terms per cut, with no
sampling overhead.

For example, the CZ gate ($k=2$) admits the
decomposition~\cite{Marshall23}
\begin{equation}
  \mathrm{CZ}
  = \frac{1}{1+i}\bigl(
      S \otimes S + i\,S^{\dagger} \otimes S^{\dagger}
    \bigr),
  \label{eq:CZ}
\end{equation}
where $S = \mathrm{diag}(1, i)$ is the phase gate.
Each term is a tensor product of single-qubit
unitaries, so cutting one CZ gate at a partition
boundary replaces it with two subcircuits that
apply $S$ or $S^{\dagger}$ to the respective
segments.
The CX gate is related by
$\mathrm{CX} = (I \otimes H)\,\mathrm{CZ}\,
(I \otimes H)$
and admits the same two-term decomposition
up to local basis changes.
When $c$~gates are cut,
the number of subcircuits grows as $k^c$ 
(exponentially with the number of cuts).

\subsection{Classical simulation}

Schr\"odinger-style (state-vector) simulation stores a state vector of size $2^{q}$ for a $q$-qubit system.
This exponential memory footprint quickly becomes the primary bottleneck.
A well-known reference point is the ``49-qubit barrier'', which motivated decomposition/slicing and storage-aware strategies to push exact simulation beyond naive memory limits \cite{Pednault17}.
Beyond exact simulation, approximate approaches deliberately trade accuracy for scale.
One direction is \emph{lossy state-vector compression}, e.g., quantization- or compression-based methods that reduce bits per amplitude while controlling the induced error/fidelity loss \cite{Huffman2024lossycompression,Wu2018amplitudeaware}.
Another direction is approximate generation of samples at a \emph{target fidelity} for structured benchmarks (e.g. Sycamore-style random circuits), which provides scalable classical baselines without computing the full exact state \cite{Pan2022sycamore}.

Tensor-network simulators \cite{markov2008simulating} can be more effective than state-vector simulation for circuits with favorable structure such as limited entanglement or small treewidth. Several large-scale approximate sampling results can also be phrased naturally in tensor-network terms.
Here we focus on the state-vector regime and quantify
when circuit cutting reduces \emph{classical wall-clock
runtime} by trading reduced subcircuit size for
cutting overhead.
\section{Execution Time Model}
\label{sec:Theory}
We model the wall-clock time of circuit cutting
on a single compute node with sequential subcircuit
execution, without gate fusion or parallelization.
Because the cutting and recombination procedure in the
classical state-vector setting exactly recovers
the full state vector of the uncut circuit
(up to machine precision),
fidelity loss does not arise and we focus exclusively
on the execution-time trade-off.

\subsection{Definitions and cost decomposition}
\label{sec:runtime_def}

To evaluate whether circuit cutting reduces the
classical simulation time, we decompose the total
cutting time as
\begin{equation}
T_{\mathrm{cut}}
= T_{\mathrm{pre}}
+ T_{\mathrm{sub}}
+ T_{\mathrm{merge}},
\label{eq:runtime_decomp}
\end{equation}
where $T_{\mathrm{pre}}$ is the preprocessing time
for generating all subcircuits and associated
coefficients,
$T_{\mathrm{sub}} = \sum_{i=1}^{N_{\mathrm{sub}}}
T_{\mathrm{SV}}^{(i)}$ is the total subcircuit
simulation time
($T_{\mathrm{SV}}^{(i)}$ is the state-vector
simulation time of the $i$-th subcircuit),
and $T_{\mathrm{merge}}$ is the merging time to reconstruct
the full state vector by combining subcircuit outcomes.
The baseline for comparison is $T_{\mathrm{orig}} = O(2^q D)$,
the wall-clock time of simulating the original circuit
without cutting.

We consider a circuit on $q$~qubits partitioned into
$n$~segments.
Let $c_i$ denote the total number of gate cuts on all
boundaries adjacent to segment~$i$,
$q_i$ the number of qubits in segment~$i$,
and $c_{\max} = \max_i c_i$.
The total number of subcircuits is
$N_{\mathrm{sub}} = \sum_{i=1}^{n} k^{c_i}$,
where $k$ is the Schmidt rank of the cut gates.

The computational cost of each phase scales
as shown in Table~\ref{tab:cost}.
Here $D$ denotes the depth of the original circuit
and $D_{\mathrm{seg},i}\,(\leq D)$ denotes the depth of
segment~$i$ after partitioning.
We write $D_{\mathrm{seg}} = \max_i D_{\mathrm{seg},i}$.
Because gate fusion is not applied,
the lower bound $D_{\mathrm{seg}} \geq D/n$
is approached when gates are distributed
uniformly across segments, giving
\begin{equation}
  \frac{D}{n} \le D_{\mathrm{seg}} \le D.
  \label{eq:dseg_bound}
\end{equation}

The preprocessing phase generates all
$N_{\mathrm{sub}}$ subcircuits by traversing
the original circuit once per subcircuit.
Each subcircuit simulation operates on a reduced
state vector of $2^{q_i}$ amplitudes, but must be
repeated for all $N_{\mathrm{sub}}$ terms.
The merging phase reconstructs the full
$2^q$-dimensional state vector by summing
$k^{c_{\max}}$ tensor products.
The memory cost of subcircuit simulation is
$O(2^{q_{\max}})$ if state vectors are discarded after
each subcircuit, or
$O(N_{\mathrm{sub}} \cdot 2^{q_{\max}})$ if all are
retained simultaneously,
where $q_{\max} = \max_i q_i$.
The pseudocodes for each phase are given in
Appendix~\ref{app:pseudocode}.

\begin{table}[htbp]
\caption{Computational cost summary.
$q$: total qubits, $n$: segments,
$c_i$: number of cuts adjacent to segment~$i$,
$c_{\max} = \max_i c_i$,
$k$: Schmidt rank of the cut gate,
$D$: original circuit depth,
$D_{\mathrm{seg}}$: maximum subcircuit depth,
$N_{\mathrm{sub}} = \sum_i k^{c_i}$: total number
of subcircuits.}
\label{tab:cost}
\begin{ruledtabular}
\begin{tabular}{lll}
Phase           & Time & Memory \\
\hline
Preprocessing   & $O(N_{\mathrm{sub}}\,D)$
                & $O(N_{\mathrm{sub}}\,D)$ \\
Subcircuit sim. & $O(N_{\mathrm{sub}}\,2^{q_{\max}}\,D_{\mathrm{seg}})$
                & $O(2^{q_{\max}})$ \\
Merging         & $O(k^{c_{\max}}\,2^q)$
                & $O(2^q)$ \\
\end{tabular}
\end{ruledtabular}
\end{table}

\subsection{Regime structure of the cutting cost}
\label{sec:regime}

The three cost components in Eq.~(\ref{eq:runtime_decomp})
exhibit qualitatively different dependences on
the circuit parameters.
To identify which component dominates $T_{\mathrm{cut}}$,
we compare the three components using their scaling
from Table~\ref{tab:cost}.

To obtain closed-form ratio expressions, we now
specialize to equal partitioning
($q_i = q/n$ for all~$i$,
$D_{\mathrm{seg},i} = D_{\mathrm{seg}}$ for all~$i$).

\subsubsection{$T_{\mathrm{pre}}$ vs $T_{\mathrm{sub}}$}
\label{sec:pre_vs_sub}

From Table~\ref{tab:cost},
\begin{equation}
  \frac{T_{\mathrm{pre}}}{T_{\mathrm{sub}}}
  = \frac{D}{2^{q/n}\,D_{\mathrm{seg}}}.
  \label{eq:pre_sub_ratio}
\end{equation}
This ratio is independent of $k$ and $c_i$.
From Eq.~(\ref{eq:dseg_bound}),
\begin{equation}
  \frac{1}{2^{q/n}}
  \le \frac{T_{\mathrm{pre}}}{T_{\mathrm{sub}}}
  \le \frac{n}{2^{q/n}}.
  \label{eq:pre_sub_bound}
\end{equation}
Since $2 \le n \le q$ and $2^{q/n} \ge 2$,
the upper bound $n/2^{q/n}$ vanishes
as $q/n$ grows,
so $T_{\mathrm{pre}} \ll T_{\mathrm{sub}}$ holds
whenever $q/n$ is sufficiently large.
For small $q/n$ (e.g.\ $q/n \lesssim 5$),
the two can be comparable.

\subsubsection{$T_{\mathrm{pre}}$ vs $T_{\mathrm{merge}}$}
\label{sec:pre_vs_merge}

From Table~\ref{tab:cost},
\begin{equation}
  \frac{T_{\mathrm{pre}}}{T_{\mathrm{merge}}}
  = \frac{N_{\mathrm{sub}}}{k^{c_{\max}}}
    \cdot \frac{D}{2^{q}}.
  \label{eq:pre_merge_ratio}
\end{equation}
The first factor satisfies
\begin{equation}
  \frac{N_{\mathrm{sub}}}{k^{c_{\max}}}
  = \sum_{i=1}^{n} k^{c_i - c_{\max}}
  \leq n,
  \label{eq:nsub_bound}
\end{equation}
since each term $k^{c_i - c_{\max}} \leq 1$.
The bound is saturated when all segments
share the same number of cuts
and decreases toward unity
as the cut distribution becomes more uneven.
In either case, this factor is $O(n)$
and does not introduce exponential dependence
on $q$ or $c$.
The ratio is therefore controlled by $D/2^{q}$.
When $2^{q} \gg D$,
$T_{\mathrm{pre}} \ll T_{\mathrm{merge}}$ holds
regardless of $n$.
For small $q$ or very deep circuits
where $D$ and $2^{q}$ are comparable,
$T_{\mathrm{pre}}$ can be comparable to
$T_{\mathrm{merge}}$.
Combined with the result of
Sec.~\ref{sec:pre_vs_sub},
$T_{\mathrm{pre}}$ is the smallest component
of $T_{\mathrm{cut}}$ whenever $q/n$
and $q$ are both sufficiently large.

\subsubsection{$T_{\mathrm{sub}}$ vs $T_{\mathrm{merge}}$}
\label{sec:sub_vs_merge}

From Table~\ref{tab:cost},
\begin{equation}
  \frac{T_{\mathrm{sub}}}{T_{\mathrm{merge}}}
  = \frac{N_{\mathrm{sub}}}{k^{c_{\max}}}
    \cdot \frac{D_{\mathrm{seg}}}{2^{q(1-1/n)}}.
  \label{eq:sub_merge_ratio}
\end{equation}
The first factor is $O(n)$
by Eq.~\eqref{eq:nsub_bound}.
The regime boundary is therefore governed by
\begin{equation}
  \frac{D_{\mathrm{seg}}}{2^{q(1-1/n)}} \approx 1.
  \label{eq:regime_boundary}
\end{equation}
When $D_{\mathrm{seg}} \gg 2^{q(1-1/n)}$
(small $q$ or large $D_{\mathrm{seg}}$),
$T_{\mathrm{sub}}$ dominates $T_{\mathrm{cut}}$
(subcircuit-dominant regime).
In the opposite limit
$D_{\mathrm{seg}} \ll 2^{q(1-1/n)}$
(large $q$ or small $D_{\mathrm{seg}}$),
$T_{\mathrm{merge}}$ dominates
(merge-dominant regime).

\subsubsection{Crossover structure}
\label{sec:crossover_structure}

The preceding ratios imply that,
at fixed depth,
increasing $q$ drives two crossovers:
$T_{\mathrm{merge}}$ overtakes $T_{\mathrm{pre}}$
and $T_{\mathrm{sub}}$.
Once $T_{\mathrm{merge}}$ overtakes both,
it is the dominant component of $T_{\mathrm{cut}}$.
Since $T_{\mathrm{merge}}$ does not benefit
from the reduced subcircuit size,
the advantage of cutting is limited
in this regime.
In Sec.~\ref{sec:threshold}, we apply this cost model
to derive quantitative threshold conditions,
which are then verified
by wall-clock measurements in
Sec.~\ref{sec:NV}.
\section{Threshold Analysis}
\label{sec:threshold}

We derive conditions on the number of cuts
under which circuit cutting reduces the classical
simulation time.
We consider a circuit on $q$~qubits with depth~$D$,
partitioned into $n$~equal segments of $q/n$~qubits each,
with $c$~cuts per boundary and
$c_{\mathrm{total}} = c(n{-}1)$ cuts in total.
We assume that gates are distributed uniformly
across segments, so that each subcircuit has depth
$D_{\mathrm{seg}} \approx D/n$.

\subsection{Cost comparison framework}
\label{sec:cost_comparison}
The total cutting time
$T_{\mathrm{cut}} = T_{\mathrm{pre}} + T_{\mathrm{sub}} + T_{\mathrm{merge}}$
has three components whose scaling is given
in Table~\ref{tab:cost}.
Of these, $T_{\mathrm{pre}}$ is never the dominant component.
The ratio analysis in Sec.~\ref{sec:regime} shows that
$T_{\mathrm{pre}}/T_{\mathrm{sub}}$ and
$T_{\mathrm{pre}}/T_{\mathrm{merge}}$
both decrease with $q$,
and the runtime breakdown in
Sec.~\ref{sec:NV_BottleneckCrossover}
confirms that $T_{\mathrm{pre}}$ never dominates
$T_{\mathrm{cut}}$ in any tested configuration.
We therefore omit $T_{\mathrm{pre}}$
from the threshold analysis.

The remaining two components involve
qualitatively different operations.
$T_{\mathrm{sub}}$ and $T_{\mathrm{orig}}$ both perform
gate-by-gate state-vector simulation
using the same simulator,
differing only in the size of the state vector
and the number of gates.
These differences are already captured
by the explicit factors $2^{q/n}$ and $D_{\mathrm{seg}}$
in Table~\ref{tab:cost},
so the order-of-magnitude comparison
$T_{\mathrm{sub}}$ vs.\ $T_{\mathrm{orig}}$
is reliable.
We derive the threshold from this pair.
In contrast,
$T_{\mathrm{merge}}$ performs vectorized tensor products
and vector additions,
which are a different type of computation
from gate-by-gate simulation.
The relative cost of these two types of operations
depends on the implementation
and is not captured by the $O(\cdot)$ scaling.
$T_{\mathrm{merge}}$ is therefore not included
in the theoretical threshold,
but is accounted for in the
empirical evaluation.

\subsection{Threshold for equal partitioning}
\label{sec:threshold_equal}

\subsubsection{Threshold under uniform gate distribution}
\label{sec:threshold_depth_free}
Under the equal-partitioning setup of this section,
comparing $T_{\mathrm{sub}}$ and $T_{\mathrm{orig}}$
from Table~\ref{tab:cost},
\begin{equation}
  N_{\mathrm{sub}} \cdot 2^{q/n} \cdot \frac{D}{n}
  \;<\; 2^{q} \cdot D,
  \label{eq:speedup_raw}
\end{equation}
which simplifies to
\begin{equation}
  \frac{N_{\mathrm{sub}}}{n}
  \;<\; 2^{q(1-1/n)}.
  \label{eq:speedup_depth_free}
\end{equation}
The depth $D$ cancels,
so this condition depends only on $q$, $n$, $c$, and~$k$.

We now set $k = 2$ (CZ gate cutting)
and derive explicit thresholds.

\paragraph{Bipartite case ($n = 2$).}
With $N_{\mathrm{sub}} = 2 \cdot 2^{c_{\mathrm{total}}}$,
Eq.~\eqref{eq:speedup_depth_free} gives
\begin{equation}
  c_{\mathrm{total}} < \frac{q}{2}.
  \label{eq:threshold_n2}
\end{equation}

\paragraph{General $n \geq 3$.}
The interior term dominates
in $N_{\mathrm{sub}}$,
giving
$N_{\mathrm{sub}} \approx (n{-}2)\cdot
2^{2c_{\mathrm{total}}/(n-1)}$.
Substituting into Eq.~\eqref{eq:speedup_depth_free}
and taking logarithms gives
\begin{equation}
  c_{\mathrm{total}} < \frac{q(n{-}1)^2}{2n}
      + \frac{(n{-}1)\log_2\!\bigl(\tfrac{n}{n-2}\bigr)}{2}.
  \label{eq:threshold_nge3}
\end{equation}
The leading slope in the
$(q,\,c_{\mathrm{total}})$ plane is $(n{-}1)^2/(2n)$,
which increases with $n$.
We denote the second term in
Eq.~\eqref{eq:threshold_nge3} by $\delta$.
For practically relevant values of $n$
(up to about $10$),
$\delta$ remains close to $1.5$.

Table~\ref{tab:threshold_values} summarizes
the threshold parameters for
$n = 2$, $3$, and $4$.
These thresholds guarantee
$T_{\mathrm{sub}} < T_{\mathrm{orig}}$
and provide a practical guideline
for selecting the number of cuts.
The additional contribution of
$T_{\mathrm{merge}}$, which grows as
$O(k^{c_{\max}}\cdot 2^q)$,
is not included in these thresholds
and can become significant at large $q$.

\begin{table}[htbp]
\caption{Subcircuit-simulation threshold
$c_{\mathrm{total}} < c_{\mathrm{total}}^{\max}$
for $k = 2$ (CZ gate cutting).
The threshold has the form
$c_{\mathrm{total}} < \mathrm{slope} \times q + \delta$.
These conditions compare $T_{\mathrm{sub}}$
with $T_{\mathrm{orig}}$ and do not include
$T_{\mathrm{merge}}$.}
\label{tab:threshold_values}
\begin{ruledtabular}
\begin{tabular}{lll}
$n$ & Slope & $\delta$ \\
\hline
$2$ & $1/2$  & $0$ \\
$3$ & $2/3$  & $\log_2 3 \approx 1.58$ \\
$4$ & $9/8$  & $3/2$ \\
\end{tabular}
\end{ruledtabular}
\end{table}

\subsubsection{Effect of non-uniform gate distribution}
\label{sec:nonuniform}

When gates are not distributed uniformly
across segments,
some subcircuits contain more gates than others,
resulting in $D_{\mathrm{seg}} > D/n$.
For general $D_{\mathrm{seg}}$,
the condition $T_{\mathrm{sub}} < T_{\mathrm{orig}}$
becomes
\begin{equation}
  \frac{N_{\mathrm{sub}}}{2^{q(1-1/n)}}
  \cdot
  \frac{D_{\mathrm{seg}}}{D}
  < 1.
  \label{eq:threshold_with_D}
\end{equation}
Since $D/n \leq D_{\mathrm{seg}} \leq D$,
the second factor $D_{\mathrm{seg}}/D$ ranges from
$1/n$ (uniform) to $1$ (worst case).
For $n = 2$, the $c_{\mathrm{total}}$ threshold
decreases by $1$ cut
compared with Eq.~\eqref{eq:threshold_n2}.
For $n \geq 3$, the reduction is
$(n{-}1)\log_2(n)/2$,
which equals $1.6$ for $n = 3$
and $3.0$ for $n = 4$.
The uniform-depth threshold
represents the most favorable case.
The reduction is modest for small $n$
but grows with $n$,
so for large $n$ the actual gate distribution
should be taken into account.

The thresholds derived in this section
compare $T_{\mathrm{sub}}$ with $T_{\mathrm{orig}}$
and do not include $T_{\mathrm{merge}}$.
The contribution of $T_{\mathrm{merge}}$
to the total cutting time is quantified empirically
in Sec.~\ref{sec:NV_BottleneckCrossover}.
\section{Evaluation}
\label{sec:NV}
We verify the threshold conditions on a classical
state-vector simulator and report a runtime breakdown.

\subsection{Experimental setup}
\label{sec:methods}

\subsubsection{Simulation environment}
All experiments were performed on a single compute node
(Apple M3 Max, 128 GB RAM) running Python 3.10.8
and Qiskit 2.2.3. State-vector simulation was carried
out using the \texttt{Statevector} class
(\texttt{qiskit.quantum\_info}). We report wall-clock
time measured on the same machine for all runs. For the
runtime breakdown measurements (Fig.~\ref{fig:bottleneck}),
memory was explicitly freed between runs to avoid
growth in the preprocessing time at large qubit counts.
Unless stated otherwise, each configuration was
measured ten times, and we report the mean values.

\subsubsection{Circuit family and partitioning}
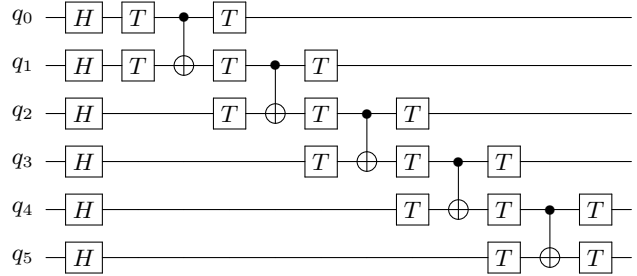
\begin{figure}[htbp]
  \centering
  \hspace{1em}
  \Qcircuit @C=0.8em @R=0.7em {
    \lstick{q_0} & \gate{H} & \gate{T} & \ctrl{1} & \gate{T} & \qw      & \qw      & \qw      & \qw      & \qw      & \qw      & \qw      & \qw      & \qw \\
    \lstick{q_1} & \gate{H} & \gate{T} & \targ    & \gate{T} & \ctrl{1} & \gate{T} & \qw      & \qw      & \qw      & \qw      & \qw      & \qw      & \qw \\
    \lstick{q_2} & \gate{H} & \qw      & \qw      & \gate{T} & \targ    & \gate{T} & \ctrl{1} & \gate{T} & \qw      & \qw      & \qw      & \qw      & \qw \\
    \lstick{q_3} & \gate{H} & \qw      & \qw      & \qw      & \qw      & \gate{T} & \targ    & \gate{T} & \ctrl{1} & \gate{T} & \qw      & \qw      & \qw \\
    \lstick{q_4} & \gate{H} & \qw      & \qw      & \qw      & \qw      & \qw      & \qw      & \gate{T} & \targ    & \gate{T} & \ctrl{1} & \gate{T} & \qw \\
    \lstick{q_5} & \gate{H} & \qw      & \qw      & \qw      & \qw      & \qw      & \qw      & \qw      & \qw      & \gate{T} & \targ    & \gate{T} & \qw
  }
  \caption{An original circuit which we can partition by one cut.
  For two or more cuts, the circuit is extended by appending additional copies of this circuit after removing the initial Hadamard gates.}
  \label{fig: original_circuit_1cut}
\end{figure}

We use a synthetic circuit family for threshold validation.
A representative instance is shown in
Fig.~\ref{fig: original_circuit_1cut}.
For two or more cuts, the circuit is extended by appending
additional copies of this circuit after removing the initial Hadamard gates.
This repeated-block structure, alternating
single-qubit and two-qubit gate layers, mirrors the
layout of practical variational and simulation circuits,
such as Ans\"atze used in
VQE~\cite{Peruzzo2014,Kandala2017},
QAOA circuits~\cite{Farhi2014},
and Trotterized Hamiltonian
simulation~\cite{Childs2021}.
In all of these, increasing the number of repeated
layers simultaneously increases both the circuit depth
and the number of two-qubit gates across partition
boundaries.

We use equal partitioning into $n$ subcircuits with
$q/n$ qubits each, since it minimizes
the simulation time, as confirmed in Appendix~\ref{app:equal_partition}.
All experiments in the remainder of this section use
equal partitioning.

\subsubsection{Runtime measurement and evaluation}
We measure $T_\mathrm{cut}$ and its three components
$T_\mathrm{pre}$, $T_\mathrm{sub}$, and $T_{\mathrm{merge}}$
as defined in Eq.~\eqref{eq:runtime_decomp}.
For each combination of $(q, c_{\mathrm{total}})$,
we evaluate the relative speedup
\begin{equation}
\Delta = (T_\mathrm{cut} - T_{\mathrm{orig}})/T_{\mathrm{orig}},
\end{equation}
where $\Delta < 0$ indicates a speedup.

\subsection{Validation of threshold conditions}
\label{sec:Validation of threshold conditions}
\begin{figure*}[htbp]
  \centering
  \includegraphics[width=\textwidth]{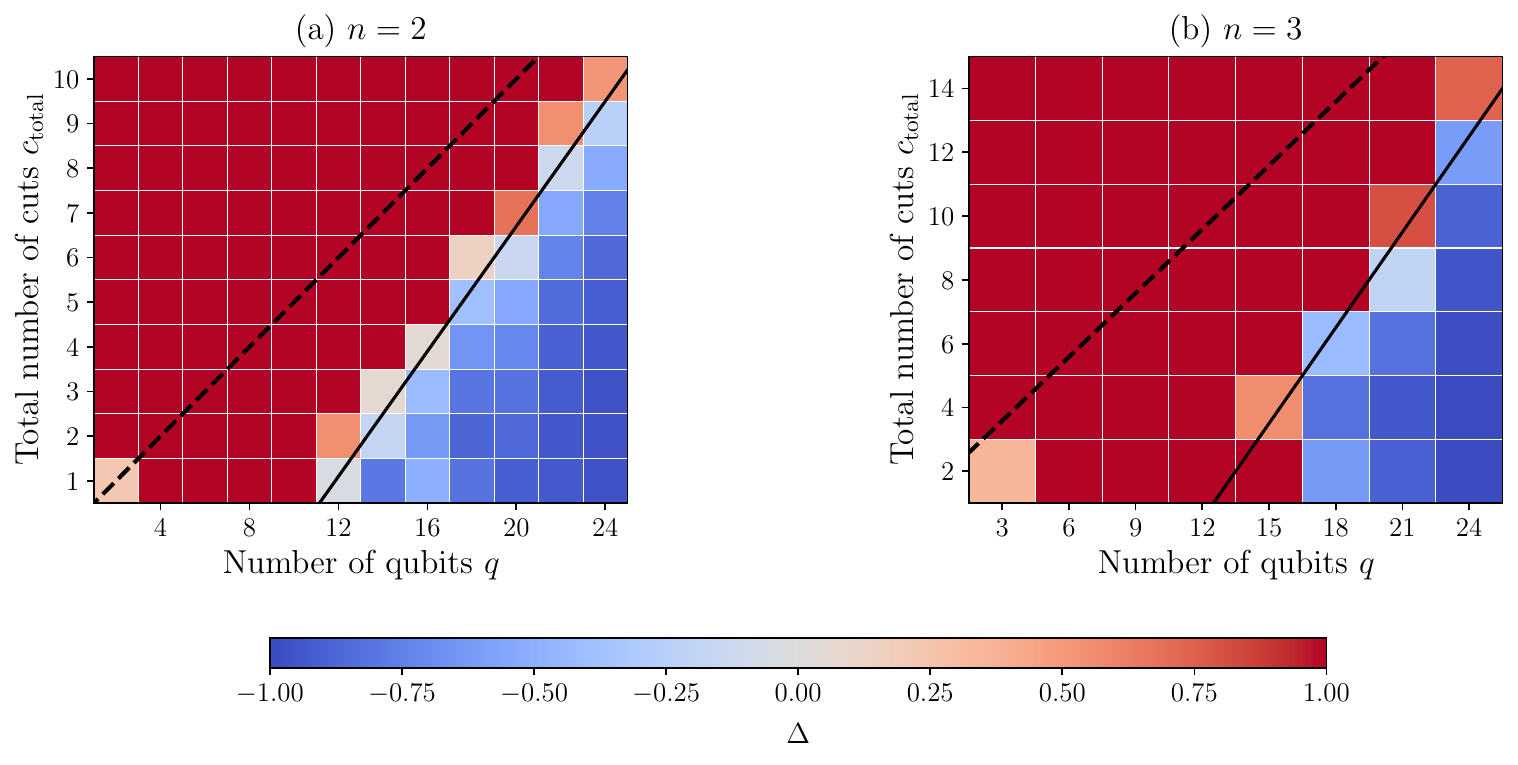}
  \caption{Measured relative speedup
  $\Delta = (T_\mathrm{cut} - T_{\mathrm{orig}})/T_{\mathrm{orig}}$
  on the $(q,\,c_{\mathrm{total}})$ plane for
  (a)~two-way ($n=2$, $q = 2,4,\ldots,24$, $c_{\mathrm{total}} = 1,\ldots,10$) and
  (b)~three-way ($n=3$, $q = 3,6,\ldots,24$, $c_{\mathrm{total}} = 2,4,\ldots,14$)
  equal partitioning.
  Blue indicates speedup ($\Delta < 0$); red indicates
  slowdown ($\Delta > 0$).
  The dashed line shows the theoretical threshold
  derived in Sec.~\ref{sec:threshold}. The solid line
  shows a linear boundary fitted to the measured data
  by a weighted SVM classifier.
  Both boundaries confirm that the speedup region
  grows linearly with~$q$, though the measured boundary
  is steeper than the theoretical prediction, as
  discussed in Sec.~\ref{sec:Validation of threshold conditions}.}
  \label{fig:heatmap_validation}
\end{figure*}

Using the experimental setup described in
Sec.~\ref{sec:methods},
the heat maps in Fig.~\ref{fig:heatmap_validation} show
the relative speedup $\Delta$ on the
$(q,\,c_{\mathrm{total}})$ plane for $n=2$ and $n=3$.
In both cases,
the boundary between the speedup region
($\Delta < 0$, blue) and the slowdown region
($\Delta > 0$, red) broadly follows the theoretical
thresholds derived in Sec.~\ref{sec:threshold}.

To quantify the measured boundary, we fit a linear
decision boundary to each panel using a weighted
support vector machine (SVM) classifier,
indicated by the solid line
in Fig.~\ref{fig:heatmap_validation}.
The SVM-fitted boundaries are
$c_{\mathrm{total}} = 0.70\,q - 7.30$ ($n=2$) and
$c_{\mathrm{total}} = 1.00\,q - 11.50$ ($n=3$),
while the theoretical thresholds,
Eqs.~\eqref{eq:threshold_n2} and \eqref{eq:threshold_nge3},
are $c_{\mathrm{total}} < q/2$ ($n=2$) and
$c_{\mathrm{total}} < 2q/3 + \log_2 3$ ($n=3$).
Comparing the two, the measured boundary has
a steeper slope and a more negative intercept.

This discrepancy arises because the theoretical
threshold compares only $T_{\mathrm{sub}}$ with
$T_{\mathrm{orig}}$, omitting $T_{\mathrm{pre}}$
and $T_{\mathrm{merge}}$.
In the benchmark circuit, depth~$D$ increases with
$c_{\mathrm{total}}$.
Since $T_{\mathrm{merge}} = O(k^{c_{\max}} \cdot 2^q)$ is independent
of depth while $T_{\mathrm{orig}} = O(2^{q} \cdot D)$,
the ratio $T_{\mathrm{merge}}/T_{\mathrm{orig}}$
decreases as $c_{\mathrm{total}}$ (and hence $D$) grows,
steepening the measured slope.
Meanwhile, $T_{\mathrm{pre}} = O(N_{\mathrm{sub}}\,D)$
grows much more slowly with~$q$ than $T_{\mathrm{sub}}$,
contributing a nearly constant offset that lowers
the intercept.
The linear form $c_{\mathrm{total}} \propto q$ is
preserved for both $n=2$ and $n=3$, confirming that
the theoretical threshold captures the correct scaling.

\subsection{Runtime breakdown and bottleneck crossovers}
\label{sec:NV_BottleneckCrossover}

\begin{figure}[htbp]
  \includegraphics[width=1.0\linewidth]{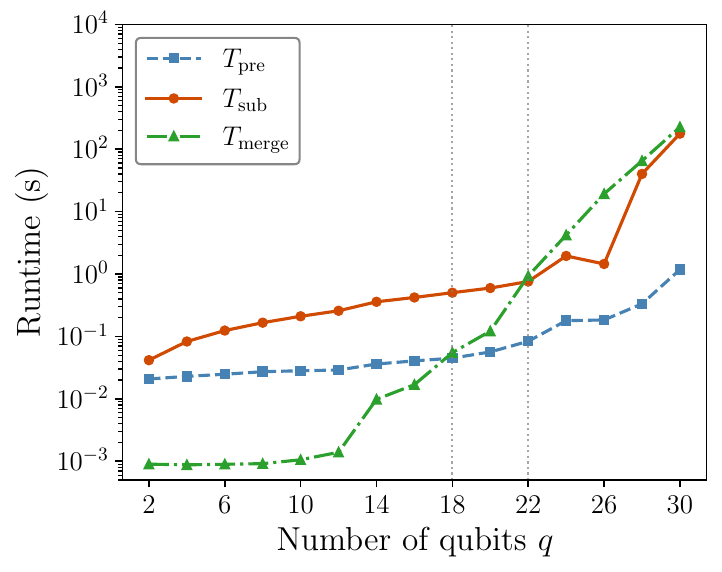}
  \caption{Runtime breakdown of two-way circuit cutting
  with $c_{\mathrm{total}}=6$ cuts, as a function of
  the number of qubits $q$.
  Three phases are shown: preprocessing ($T_\mathrm{pre}$),
  subcircuit simulation ($T_{\mathrm{sub}}$), and merging
  ($T_{\mathrm{merge}}$).
  Vertical dotted lines mark the two crossover points:
  around $q \approx 18$, where merging is no longer a
  minor overhead relative to preprocessing, and around
  $q \approx 22$, where merging overtakes subcircuit
  simulation.}
  \label{fig:bottleneck}
\end{figure}

To identify the dominant cost factors of circuit cutting, we decompose
the total cutting time into three phases: preprocessing
($T_\mathrm{pre}$), subcircuit simulation
($T_{\mathrm{sub}}$), and merging ($T_{\mathrm{merge}}$).
We report these three components as a function of $q$
in Fig.~\ref{fig:bottleneck} for two-way cutting with
$c_{\mathrm{total}}=6$.
Each data point represents the mean of ten independent runs.

For small circuits ($q \lesssim 16$), preprocessing and subcircuit
simulation account for most of the total cutting time, while the merging cost is
negligible ($< 10^{-1}$~s).
However, $T_{\mathrm{merge}}$ grows most steeply with
$q$, because the merging step reconstructs the full
$2^q$-dimensional state vector from the subcircuit
results.
This leads to two crossover points:

\begin{itemize}
  \item \textbf{First crossover ($q \approx 18$):}
    $T_{\mathrm{merge}}$ surpasses $T_\mathrm{pre}$,
    indicating that merging is no longer a minor
    overhead relative to preprocessing.
  \item \textbf{Second crossover ($q \approx 22$):}
    $T_{\mathrm{merge}}$ overtakes
    $T_{\mathrm{sub}}$, becoming the dominant cost among the three phases.
    For $q = 30$, merging accounts for approximately
    56\% of the total cutting time, while subcircuit
    simulation accounts for 43\% and preprocessing is
    negligible ($< 1\%$).
\end{itemize}

\noindent
The same qualitative trend is observed for other cut counts
in Fig.~\ref{fig:crossover_panels} in
Appendix~\ref{sec:SupplementaryFigures}.

These crossovers suggest different optimization targets
depending on circuit size.
For small circuits ($q \lesssim 18$ in our setup),
the total cutting time is governed by preprocessing and
subcircuit simulation, and standard optimization
strategies (e.g.\ gate fusion, parallel subcircuit
simulation) are most effective.
For larger circuits ($q \gtrsim 22$ in our setup),
optimizing the merging step, for example by employing
sparse or compressed state-vector
representations~\cite{Huffman2024lossycompression,Wu2018amplitudeaware},
yields the greatest reduction in total cutting time.

\subsection{Depth sweep and runtime breakdown}
\label{sec:depthsweep}

The thresholds in
Eqs.~\eqref{eq:threshold_n2} and~\eqref{eq:threshold_nge3}
do not contain the circuit depth~$D$, because $D$
cancels in the comparison of $T_{\mathrm{sub}}$ with
$T_{\mathrm{orig}}$ under uniform gate distribution.
The speedup boundary is therefore expected to be
stable across depths.
The depth sweep below tests this expectation
empirically.

In practical state-vector simulation, however, the
wall-clock time also depends on the circuit depth
(or, equivalently, the number of gate applications).
To quantify this effect and to assess the robustness
of the threshold boundary, we perform a synthetic
depth sweep by appending $10^p$ layers of
single-qubit gates (alternating $T$ and $X$) to the
top and bottom qubits of a baseline circuit, while
keeping $(q, c_{\mathrm{total}})$ fixed.
The circuit diagram is given in
Fig.~\ref{fig:depth_circuit} in
Appendix~\ref{sec:SupplementaryFigures}.

\begin{figure}[htbp]
  \centering
  \includegraphics[width=\columnwidth]{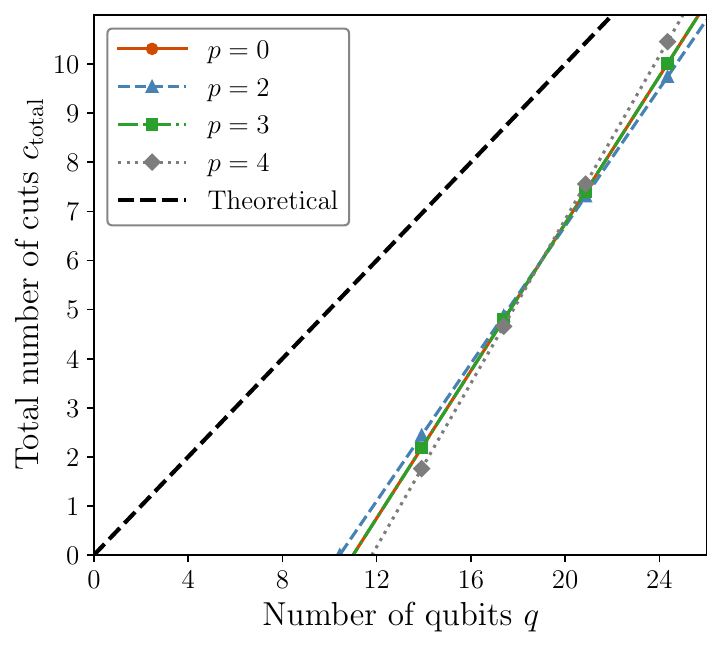}
  \caption{Speedup boundary on the $(q,\,c_{\mathrm{total}})$ plane for
  two-way partitioning ($n=2$) at four depth
  configurations ($p = 0, 2, 3, 4$, corresponding to
  $D \propto 1, 10^2, 10^3, 10^4$ additional gate layers).
  Each colored line shows the $\Delta = 0$ boundary
  obtained by a weighted SVM fit at the corresponding
  depth. Cutting is beneficial below each line.
  The dashed black line shows the theoretical threshold
  derived in Sec.~\ref{sec:threshold}.
  The boundaries remain close to one another across all
  tested depths, confirming that the qubit-count-based
  threshold provides a reasonable first-order guideline
  even for circuits of varying depth.}
  \label{fig:depth_boundary}
\end{figure}

We extract the $\Delta=0$ boundary on the
$(q, c_{\mathrm{total}})$ plane for each depth
configuration and overlay them in
Fig.~\ref{fig:depth_boundary}.
Specifically, for each depth~$D$ we compute
$\Delta(q, c_{\mathrm{total}}; D)
= (T_\mathrm{cut}(q, c_{\mathrm{total}}; D)
- T_{\mathrm{orig}}(q; D)) / T_{\mathrm{orig}}(q; D)$
and fit a linear decision boundary using a weighted
SVM classifier.
Cutting is beneficial below each fitted line.
The resulting boundaries remain close to one another
across all tested depths ($p=0,2,3,4$), confirming
that the speedup--slowdown transition is
qualitatively stable.
Increasing depth introduces two competing effects:
$T_{\mathrm{orig}}$ grows with depth, which tends to
make the cutting overhead relatively smaller, while
$T_{\mathrm{sub}}$ also grows with depth, partially
offsetting this advantage.
These effects largely cancel, so the speedup boundary
remains approximately stable.
This stability refers to the boundary as a whole.
At the level of individual configurations, depth can
still tip the balance in favor of cutting, as detailed
in Fig.~\ref{fig:runtime_depth} in
Appendix~\ref{sec:SupplementaryFigures}.

\begin{figure}[hbtp]
  \centering
  \includegraphics[width=\linewidth]{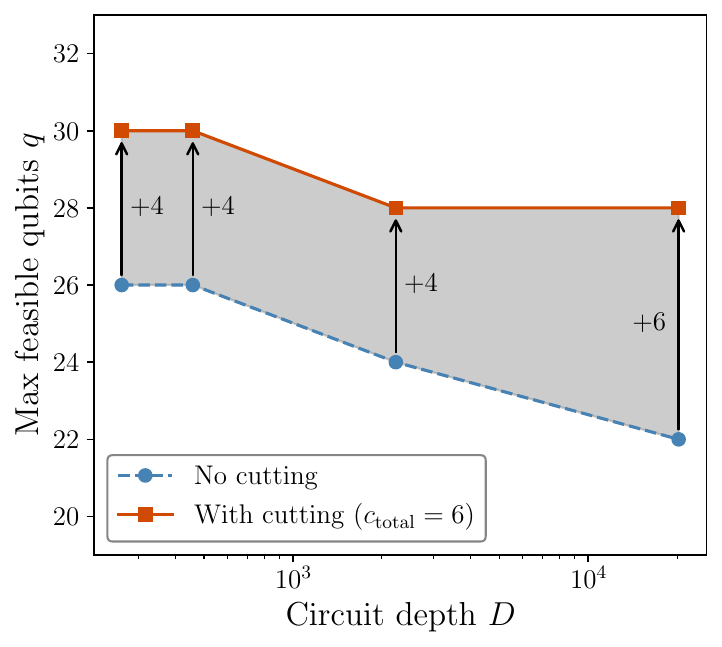}
  \caption{Maximum feasible qubit count within a 10-minute
  wall-clock budget, as a function of circuit depth~$D$.
  The dashed blue line shows the original (uncut) simulation
  and the solid orange line shows two-way cutting with
  $c_{\mathrm{total}}=6$.
  The shaded region highlights the improvement from cutting.
  Annotated arrows indicate the qubit-count gain at each
  depth: 4 qubits for $p=0,2,3$ and 6 qubits for $p=4$,
  extending the feasible range from 22--26 (no cutting)
  to 28--30 (with cutting).}
  \label{fig:max_feasible_q}
\end{figure}

To provide an operational guideline, we also estimate the maximum feasible
qubit count within a fixed wall-clock budget for both the original simulation
and two-way cutting, as shown in Fig.~\ref{fig:max_feasible_q}.
We set the time limit to 10~minutes per simulation run.
This time limit was chosen because
(i)~it is practical for iterative circuit design and verification workflows, and
(ii)~all reported results are determined entirely from measured simulation times
without any extrapolation. Every configuration either completed within the
limit or timed out, so the feasibility boundary is exact.
Each configuration was executed three independent times.
The results were consistent across runs, and the figure reports the mean values.
Under this 10-minute budget,
the original (uncut) simulation reaches at most
22--26 qubits across all depth configurations
($p = 0, 2, 3, 4$), while two-way cutting extends
the feasible range to 28--30 qubits, a gain of
4 to 6 qubits, as indicated by the annotated arrows
in Fig.~\ref{fig:max_feasible_q}.
The shaded region highlights the gain from cutting
across all depths.

Moreover, the gain increases from 4 to 6 qubits
as depth grows from $p=0$ to $p=4$, consistent with
the observation that $T_\mathrm{cut}$ grows more slowly
with depth than $T_{\mathrm{orig}}$, as shown in
Fig.~\ref{fig:runtime_depth} in
Appendix~\ref{sec:SupplementaryFigures}.

\subsection{Scope and limitations}
\label{sec:scope-and-limitations}

The experiments in this section confirm 
four qualitative trends within the tested range 
($q \leq 24$ for the threshold boundary, 
$q \leq 30$ for the runtime breakdown): 
(i) the speedup boundary is linear in $q$, 
(ii) $T_\mathrm{merge}$ becomes the dominant cost component 
at large qubit counts, 
(iii) the boundary is approximately stable 
across a wide range of circuit depths, and 
(iv) circuit cutting extends the feasible simulation range 
under a fixed wall-clock budget. 
All four trends originate from the exponential scaling 
of state-vector simulation and are expected to persist 
across different platforms and simulator implementations. 
The quantitative values, however,
including the SVM-fitted slopes and intercepts,
the bottleneck crossover points,
and the feasibility gain,
are specific to our single-node environment,
benchmark circuit family, and simulator implementation,
and should be recalibrated for each target platform.
The runtime decomposition methodology
and the threshold analysis on the $(q, c_{\mathrm{total}})$ plane
are directly applicable to other environments
and provide the framework for such recalibration.

The execution times are discussed in terms of circuit
depth~$D$ following the convention in simulation
complexity studies.
For sequential gate-application simulators such as
Qiskit \texttt{Statevector}, the wall-clock time
scales with gate count~$G$ rather than with~$D$, but the two coincide
under the uniform gate distribution assumed in the
threshold analysis and in the numerical experiments.
\section{Conclusion}
\label{sec:conclusion}
We studied the practical conditions under which circuit
cutting accelerates classical state-vector simulation with
full-state reconstruction.
Our main contributions are summarized as follows.

(i) Decomposing the total cutting time into preprocessing,
subcircuit simulation, and merging revealed two
bottleneck crossovers: merging surpasses preprocessing
and then surpasses subcircuit simulation.
In our implementation, these crossovers occur around
$q \approx 18$ and $q \approx 22$, respectively.

(ii) From this cost model we derived conditions on
the maximum cut count $c_{\mathrm{total}}$ for which
cutting yields a net speedup.
The threshold takes the form
$c_{\mathrm{total}} \propto q$ and is approximately
depth-independent.
State-vector experiments up to 24~qubits confirm
the predicted threshold.

(iii) As a practical illustration, we estimated the
maximum feasible qubit count under a 10-minute
wall-clock budget. Two-way cutting extends this
limit by 4 to 6~qubits compared to simulation
without cutting, with the gain increasing at larger
depths.

These results also offer a perspective on the well-known
49-qubit barrier for state-vector
simulation~\cite{Pednault17}.
Circuit cutting does not unconditionally break this
barrier, because the exponential overhead in cut count
implies that only circuits requiring a small number of
cuts can achieve a net speedup.
However, the threshold conditions provide a practical
pre-screening criterion: given a circuit's
$(q,\,c_{\mathrm{total}})$ parameters, one can determine
in advance whether cutting is likely to yield a net
speedup, without committing to the full simulation.

The experimental validation covers $q \leq 24$ for the
threshold boundary and $q \leq 30$ for the runtime
breakdown.
Beyond this range, the threshold predictions rely on
extrapolation of the order-of-magnitude scaling, and
memory constraints become the practical bottleneck
for single-node state-vector simulation.
The quantitative crossover points and feasibility gains
reported here are specific to our single-node environment
and should be recalibrated for each target platform.
The functional form of the threshold and the regime
structure are expected to
persist across different environments.

This work focuses on full-state reconstruction, where
the $O(2^q)$ merging cost is unavoidable.
When only expectation values are needed, the merging
phase can be replaced by classical summation of
subcircuit expectation values, which eliminates the
$O(2^q)$ cost and changes the regime structure.
Extending the present framework to that setting is a
natural next step.

Several further directions remain open.
First, the threshold model can be refined by extending
to asymmetric and topology-aware partitioning, and
the effective threshold can be further relaxed by
parallel subcircuit execution and more efficient
merging algorithms.
Second, applying this framework to application-specific
circuits, such as those arising in quantum chemistry
and combinatorial optimization, would enable practical
pre-screening of whether circuit cutting yields a net
speedup in realistic workflows.
Third, incorporating quantum execution costs, including
shot count and state tomography overhead, would extend
the analysis to a quantum--classical crossover regime.
This requires a separate cost model for the quantum side and is left for future work.

\begin{acknowledgments}
This paper is based on results obtained from a project,
JPNP23003, commissioned by the New Energy and Industrial Technology Development Organization (NEDO).
The authors would like to thank Junya Nakamura, Tsuyoshi Kitano, Hana Ebi, Hideaki Kawaguchi,  Takaharu Yoshida, Hiroki Kuji and Shigetora Miyashita for their fruitful discussions. 
TS is supported by 
JST Grant Number JPMJPF2221.
\end{acknowledgments}

\appendix
\section{SUPPLEMENTARY FIGURES}
\label{sec:SupplementaryFigures}

This appendix collects supporting visualizations
that provide additional evidence for the claims
in the main text. It covers equal-partitioning
optimality, runtime breakdown across cut counts,
and depth-sweep details.

\subsection{Optimality of equal partitioning}
\label{app:equal_partition}
\begin{figure}[htbp]
  \centering
  \includegraphics[width=\linewidth]{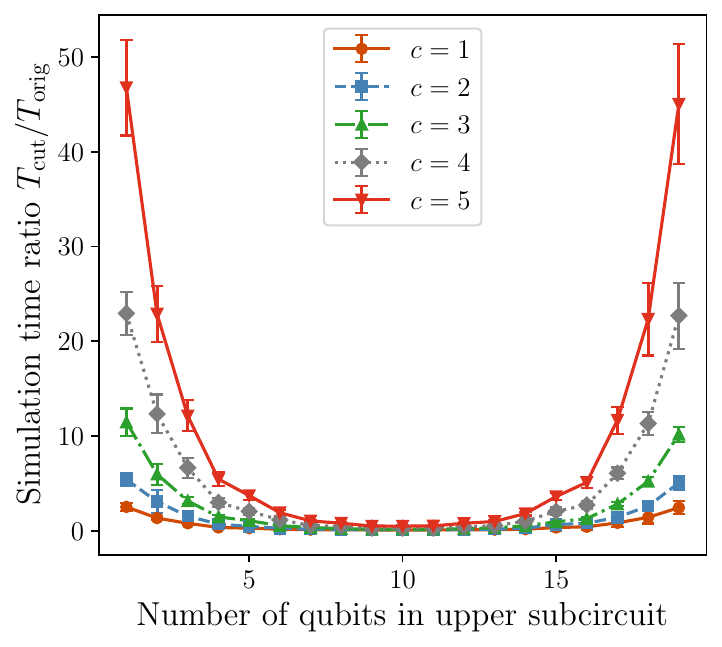}
  \caption{Simulation time ratio
  $T_{\mathrm{cut}}/T_{\mathrm{orig}}$ as a function of
  the number of qubits in the upper subcircuit for a
  20-qubit circuit partitioned into two subcircuits with
  $c = 1, \ldots, 5$ cuts. The minimum occurs at
  $q/2 = 10$ for all cut counts, confirming that equal
  partitioning minimizes the total simulation time.}
  \label{fig:equal_partition}
\end{figure}

We verify that equal partitioning minimizes the
total simulation time.
A 20-qubit circuit is partitioned into two subcircuits
by varying the number of qubits in the upper subcircuit
from 1 to 19. For each configuration, the simulation
time is measured 20 times and the mean and standard
deviation are computed for 1 to 5 cuts.
As shown in Fig.~\ref{fig:equal_partition},
the simulation time ratio is minimized when the circuit
is divided into equal halves ($q/2 = 10$ qubits per
subcircuit).

\subsection{Runtime breakdown across cut counts}
\label{app:crossover_panels}

\begin{figure*}[htbp]
  \centering
  \includegraphics[width=\textwidth]{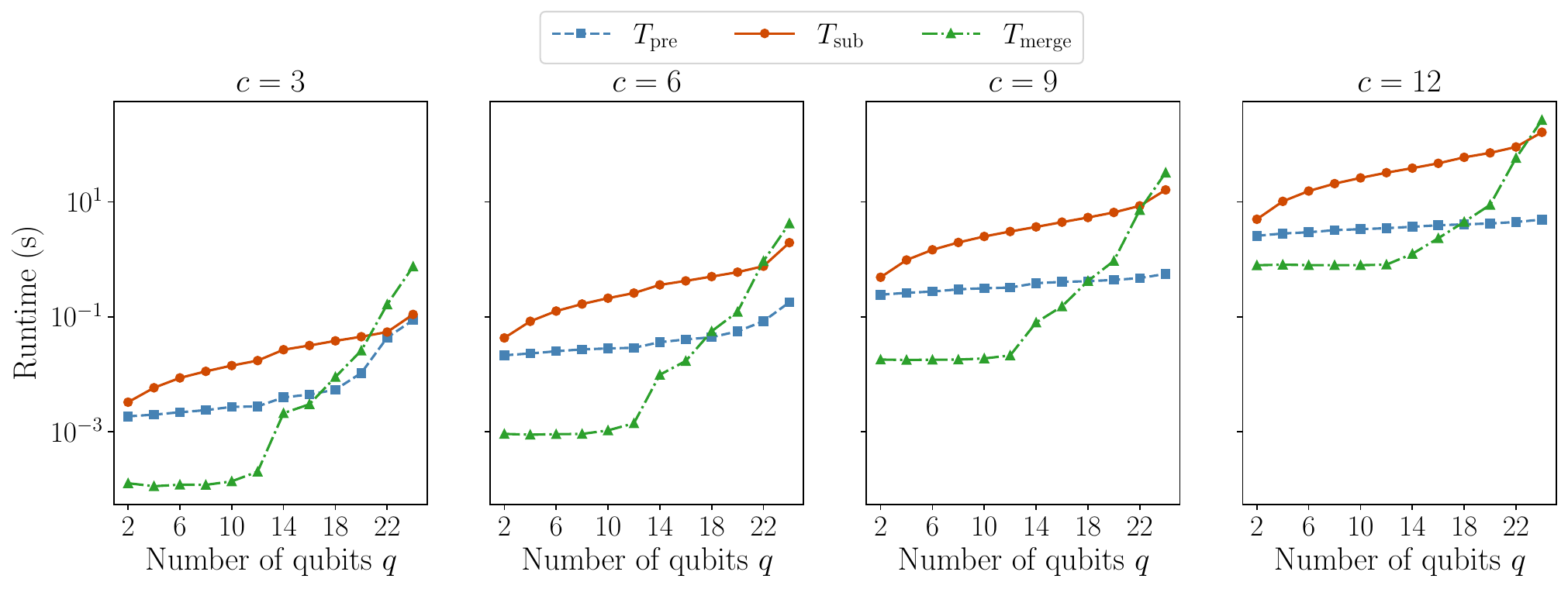}
  \caption{Runtime breakdown of two-way cutting ($n=2$)
  for four representative cut counts
  $c = 3,\,6,\,9,\,12$. Each panel shows the three cost
  components: preprocessing ($T_\mathrm{pre}$),
  subcircuit simulation
  ($T_{\mathrm{sub}}$), and merging
  ($T_{\mathrm{merge}}$) as a function of the
  number of qubits~$q$. The same qualitative trend as in
  Fig.~\ref{fig:bottleneck} is observed across all cut
  counts: $T_{\mathrm{merge}}$ grows most steeply with~$q$,
  producing two crossover points similar to those
  identified in Sec.~\ref{sec:NV_BottleneckCrossover}.
  The $c=6$ panel corresponds to
  Fig.~\ref{fig:bottleneck} in the main text and is
  included for ease of comparison.}
  \label{fig:crossover_panels}
\end{figure*}

The runtime breakdown for two-way cutting at four
representative cut counts ($c = 3, 6, 9, 12$) is
shown in Fig.~\ref{fig:crossover_panels}, including
the $c = 6$ case in Fig.~\ref{fig:bottleneck}.
The same qualitative trend as in Fig.~\ref{fig:bottleneck}
is observed across all cut counts: $T_{\mathrm{merge}}$
grows most steeply with $q$, producing two crossover
points similar to those identified in
Sec.~\ref{sec:NV_BottleneckCrossover}.
As $c$ increases, the crossover points shift slightly
toward larger $q$, but the overall structure of the
runtime breakdown remains unchanged.

\subsection{Depth sweep: circuit and runtime breakdown}
\label{app:depth_sweep}
\begin{figure}[htbp]
  \centering
  \resizebox{0.95\columnwidth}{!}{
    \begin{tikzpicture}[font=\huge]
      \node (circuit) {
        \Qcircuit @C=0.3em @R=0.3em {
          \lstick{q_0} & \gate{H} & \gate{T} & \gate{X} & \push{\cdots} & \gate{T} & \gate{X} & \gate{T} & \ctrl{1} & \gate{T} & \qw      & \qw      & \qw      & \qw      & \qw      & \qw      & \qw      & \qw      & \qw      & \qw      & \qw      & \qw      & \qw      & \qw \\
          \lstick{q_1} & \gate{H} & \qw      & \qw      & \qw           & \qw      & \qw      & \gate{T} & \targ    & \gate{T} & \ctrl{1} & \gate{T} & \qw      & \qw      & \qw      & \qw      & \qw      & \qw      & \qw      & \qw      & \qw      & \qw      & \qw      & \qw \\
          \lstick{q_2} & \gate{H} & \qw      & \qw      & \qw           & \qw      & \qw      & \qw      & \qw      & \gate{T} & \targ    & \gate{T} & \ctrl{1} & \gate{T} & \qw      & \qw      & \qw      & \qw      & \qw      & \qw      & \qw      & \qw      & \qw      & \qw \\
          \lstick{q_3} & \gate{H} & \qw      & \qw      & \qw           & \qw      & \qw      & \qw      & \qw      & \qw      & \qw      & \gate{T} & \targ    & \gate{T} & \ctrl{1} & \gate{T} & \qw      & \qw      & \qw      & \qw      & \qw      & \qw      & \qw      & \qw \\
          \lstick{q_4} & \gate{H} & \qw      & \qw      & \qw           & \qw      & \qw      & \qw      & \qw      & \qw      & \qw      & \qw      & \qw      & \gate{T} & \targ    & \gate{T} & \ctrl{1} & \gate{T} & \qw      & \qw      & \qw      & \qw      & \qw      & \qw \\
          \lstick{q_5} & \gate{H} & \qw      & \qw      & \qw           & \qw      & \qw      & \qw      & \qw      & \qw      & \qw      & \qw      & \qw      & \qw      & \qw      & \gate{T} & \targ    & \gate{T} & \gate{T} & \gate{X} & \push{\cdots} & \gate{T} & \gate{X} & \qw
        }
      };
      \draw [decorate, decoration={brace, amplitude=5pt}]
        ([shift={(-1.25\textwidth, 0.15cm)}]circuit.east |- circuit.north) -- ([shift={(-0.85\textwidth, 0.15cm)}]circuit.east |- circuit.north)
        node [midway, above=5pt, font=\Huge] {$10^p$};
      \draw [decorate, decoration={brace, amplitude=5pt, mirror}]
        ([shift={(-0.32\textwidth, -0.15cm)}]circuit.east |- circuit.south) -- ([shift={(-0.02\textwidth, -0.15cm)}]circuit.east |- circuit.south)
        node [midway, below=6pt, font=\Huge] {$10^p$};
    \end{tikzpicture}
  }
  \caption{Depth-augmented circuit used in the depth sweep.
  $10^p$ layers of alternating $T$ and $X$ gates are appended
  to the top and bottom qubits of the baseline circuit.
  The number of qubits $q$ and cuts $c$ are kept fixed
  while only the circuit depth is varied.}
  \label{fig:depth_circuit}
\end{figure}
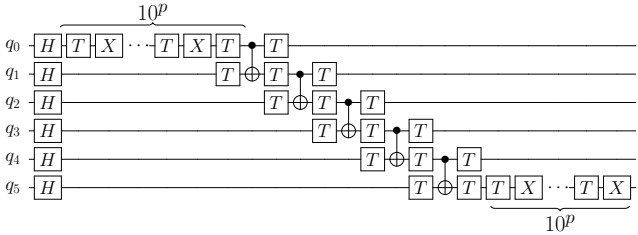

The depth-augmented circuit used in the depth sweep
of Sec.~\ref{sec:depthsweep} is shown in
Fig.~\ref{fig:depth_circuit}.
$10^p$ layers of alternating $T$ and $X$ gates are
appended to the top and bottom qubits of the baseline
circuit, while the number of qubits $q$ and cuts $c$
are kept fixed.

\begin{figure}[hbtp]
  \centering
  \includegraphics[width=\linewidth]{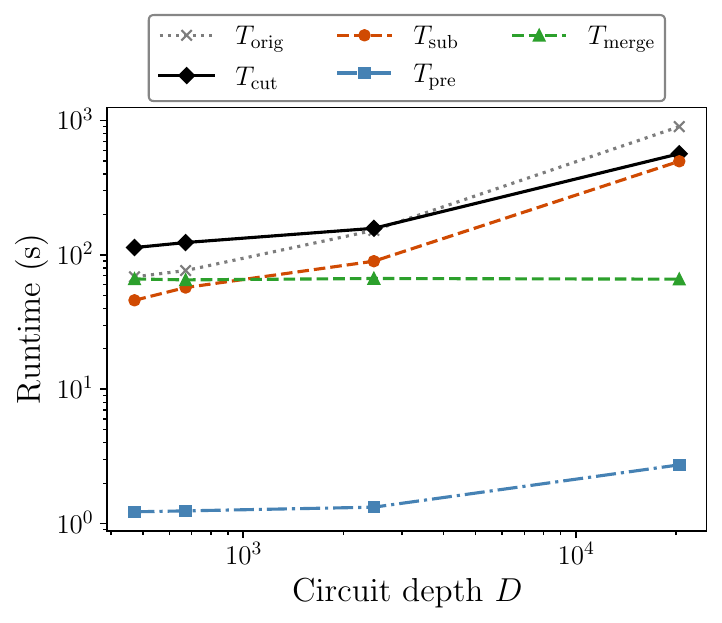}
  \caption{Runtime breakdown as a function of circuit
  depth $D$ for a 24-qubit circuit with 10 cuts, divided
  into 2 segments. The total cutting time
  ($T_\mathrm{cut}$) is primarily composed of the
  subcircuit simulation time
  ($T_\mathrm{sub}$), and both exhibit a
  similar scaling trend with depth. In contrast, the
  merging overhead ($T_{\mathrm{merge}}$) remains
  approximately constant, while the preprocessing cost
  ($T_\mathrm{pre}$) shows a weaker dependence on depth.
  Around $D \approx 750$, the dominant bottleneck within
  $T_\mathrm{cut}$ shifts from $T_{\mathrm{merge}}$ to
  $T_\mathrm{sub}$. The original simulation time
  without cutting ($T_\mathrm{orig}$) grows faster than
  $T_\mathrm{cut}$ and overtakes it around
  $D \approx 3000$,
  indicating that circuit cutting can become advantageous
  at sufficient depth, even for configurations that do not
  yield speedup at small depth.}
  \label{fig:runtime_depth}
\end{figure}

We report the runtime breakdown
$\{T_\mathrm{pre},\,T_\mathrm{sub},\,
T_{\mathrm{merge}}\}$ as a function of circuit depth $D$
for a representative configuration ($q = 24$, $c = 10$,
two-way partitioning) in Fig.~\ref{fig:runtime_depth}.
Among the three components, $T_\mathrm{pre}$ and
$T_\mathrm{sub}$ grow approximately linearly with $D$,
whereas $T_{\mathrm{merge}}$ is approximately
depth-independent.
As a result, the dominant component within
$T_\mathrm{cut}$ shifts from $T_{\mathrm{merge}}$ to
$T_\mathrm{sub}$ around $D \approx 750$.
Because $T_{\mathrm{merge}}$ does not grow with depth,
$T_\mathrm{cut}$ grows more slowly than
$T_\mathrm{orig}$, and $T_\mathrm{orig}$ overtakes
$T_\mathrm{cut}$ around $D \approx 3000$.

\FloatBarrier
\section{PSEUDOCODE AND COMPUTATIONAL COST}
\label{app:pseudocode}

This appendix provides the pseudocodes underlying the
cost expressions in Table~\ref{tab:cost}.
All pseudocodes assume CZ gate cutting with $k=2$.
In the preprocessing phase (Algorithm~1), each CX gate
at a partition boundary is first converted to CZ via
$\mathrm{CX} = (I \otimes H)\,\mathrm{CZ}\,(I \otimes H)$
and then decomposed.

\begin{algorithm}[H]
\caption{Preprocessing}
\label{alg:cut}
\begin{algorithmic}[1]
\Require Original circuit $\mathcal{C}$ on $q$ qubits,
         partition $\{G_1,\dots,G_n\}$ of qubits
\Ensure  Subcircuits $\{S_i^{(r)}\}$
         for segment $i=1,\dots,n$ and
         decomposition index $r=1,\dots,2^{c_i}$,
         where $c_i$ is the total number of cuts
         on all boundaries adjacent to segment~$i$.
         Merge index table $\mathcal{M}$.
\Statex
\State Scan $\mathcal{C}$ and identify all CX gates
       that straddle partition boundaries
\For{each segment $i = 1,\dots,n$}
    \State Split $\mathcal{C}$ at each boundary gate into
           gate-list fragments
           $F_i^{(0)},F_i^{(1)},\dots,F_i^{(c_i)}$
           restricted to qubits in $G_i$
    \For{each combination $\mathbf{b}\in\{0,1\}^{c_i}$}
        \State $S_i^{(\mathbf{b})} \gets F_i^{(0)}$
        \For{$j = 1,\dots,c_i$}
            \If{$b_j = 0$}
                \State Append $HSH$ (or $S$) gate to cut qubit
            \Else
                \State Append $HS^\dagger H$ (or $S^\dagger$) gate to cut qubit
            \EndIf
            \State Append fragment $F_i^{(j)}$
        \EndFor
    \EndFor
\EndFor
\State Build merge index table $\mathcal{M}$ mapping each global
       combination index to the per-segment indices $\mathbf{b}$
\end{algorithmic}
\textbf{Complexity.}
Time: $O(N_{\mathrm{sub}} \cdot D)$,
where $N_{\mathrm{sub}} = \sum_{i=1}^{n} 2^{c_i}$
is the total number of subcircuits.
When every boundary has the same number of cuts~$c$,
edge segments have $c_i = c$,
interior segments have $c_i = 2c$,
and $N_{\mathrm{sub}} = 2 \cdot 2^{c}$ for $n=2$.
Memory: $O(N_{\mathrm{sub}} \cdot D)$.
\end{algorithm}

\begin{algorithm}[H]
\caption{Subcircuit simulation}
\label{alg:simulate}
\begin{algorithmic}[1]
\Require Subcircuits $\{S_i^{(r)}\}$ from Algorithm~\ref{alg:cut}
\Ensure  State vectors $\{|\psi_i^{(r)}\rangle\}$,
         each of dimension $2^{q_i}$
\Statex
\For{each segment $i = 1,\dots,n$}
    \For{each decomposition index $r = 1,\dots,2^{c_i}$}
        \State $|\psi_i^{(r)}\rangle
               \gets \text{StateVectorSimulate}(S_i^{(r)})$
               \Comment{$O(2^{q_i}\!D_{\mathrm{seg},i})$}
    \EndFor
\EndFor
\end{algorithmic}
\textbf{Complexity.}
Time: $O\!\left(\sum_{i=1}^{n} 2^{c_i} \cdot 2^{q_i} \cdot D_{\mathrm{seg},i}\right)$,
where $q_i$ is the number of qubits and
$D_{\mathrm{seg},i}$ is the depth of segment~$i$
(e.g.\ $O(2 \cdot 2^{c} \cdot 2^{q/2} \cdot D_{\mathrm{seg}})$
for equal bipartition with $c$~cuts).
Memory: $O(2^{q_{\max}})$ per subcircuit, up to
$O(N_{\mathrm{sub}} \cdot 2^{q_{\max}})$ if all state vectors
are retained for merging,
where $q_{\max} = \max_i q_i$.
\end{algorithm}

\begin{algorithm}[H]
\caption{Merging}
\label{alg:merge}
\begin{algorithmic}[1]
\Require State vectors $\{|\psi_i^{(r)}\rangle\}$
         from Algorithm~\ref{alg:simulate},
         merge index table $\mathcal{M}$
\Ensure  Reconstructed full state vector
         $|\Psi\rangle$ of dimension $2^q$
\Statex
\State Compute $\{\alpha_m\}$ from CZ decomposition
       \Comment{$O(2^{c_{\max}})$}
\State $|\Psi\rangle \gets \mathbf{0}_{2^q}$
\For{each combination $m = 1,\dots,2^{c_{\max}}$}
    \State Look up per-segment indices
           $(r_1,\dots,r_n) \gets \mathcal{M}[m]$
    \State $|\phi_m\rangle \gets
           |\psi_1^{(r_1)}\rangle \otimes \cdots
           \otimes |\psi_n^{(r_n)}\rangle$
           \Comment{$O(2^q)$}
    \State $|\Psi\rangle \gets |\Psi\rangle
           + \alpha_m\,|\phi_m\rangle$
           \Comment{$O(2^q)$}
\EndFor
\end{algorithmic}
\textbf{Complexity.}
Time: $O(2^{c_{\max}} \cdot 2^{q})$,
where $c_{\max} = \max_i c_i$
($2^{c_{\max}}$ tensor products, each producing a
$2^{q}$-dimensional vector;
e.g.\ $O(2^{c} \cdot 2^{q})$ for $n=2$).
Memory: $O(2^{q})$
(one accumulator vector $|\Psi\rangle$ and one temporary
vector $|\varphi_m\rangle$).
\end{algorithm}

\bibliography{references}

\end{document}